\documentclass[aps,twocolumn,prb]{revtex4}
\usepackage{amssymb}

\input{tcilatex}

\begin{document}

\title{MASSIVE SPIN COLLECTIVE MODE IN QUANTUM HALL FERROMAGNET}
\author{T.Maniv$^{1,2}$ , Yu. A. Bychkov$^{1,3}$, I.D. Vagner$^{1,4}$ \\
$^{1}$Grenoble High Magnetic Field Laboratory,Max-Planck-\\
Institute fur Festkorperforschung and CNRS, Grenoble, France.\\
$^{2}$Chemistry Department, Technion-Israel Institute of Technology, Haifa
32000, Israel\\
$^{3}$L.D.Landau Institute for Theoretical Physics, Kosygina 2, Moscow,Russia%
\\
$^{4}$RCQCE at the Department of Communication Engineering, Holon Academic
Institute of Technology,52 Golomb str., 58102, Holon, Israel}
\date{\today{}}

\begin{abstract}
It is shown that the collective spin rotation of a single Skyrmion in
quantum Hall ferromagnet can be regarded as precession of the entire spin
texture in the external magnetic field, with an effective moment of inertia
which becomes infinite in the zero g-factor limit. This low-lying spin
excitation may dramatically enhance the nuclear spin relaxation rate via the
hyperfine interaction in the quantum well slightly away from filling factor $%
\nu =1$.

PACS numbers: 73.43.-f, 73.43.Lp, 71.70.Di, 76.60.-k, 12.39.Dc
\end{abstract}

\maketitle

Nuclear spin dynamics in semiconducting heterojunctions under the conditions
of the odd integer Quantum Hall (QH) effect is strongly influenced by the
two-dimensional (2D) electron gas in the quantum well through the hyperfine
interaction with the electronic spins. This effect was first demonstrated in
a set of experiments, reported in \cite{Barret95,Tycko95}, where the Knight
shift, $K_{S}$, and the $^{71}$Ga spin lattice relaxation time $T_{1}$ in
GaAs multiple quantum well (MQW) structure under perpendicular magnetic
field were detected by means of the optically pumped NMR (OPNMR) technique. $%
K_{S}$ was found to reduce dramatically as the Landau level filling factor
was shifted slightly away from $\nu =1$, indicating that the injection of a
single charge into the 2D electron system is followed by reversal of many
electronic spins. In the same interval of the filling factor $T_{1}$ was
found to drop by several orders of magnitude. Both effects are considered as
strong evidence for the creation of skyrmionic spin texture \cite{Sondhi93}, 
\cite{BMV96},\cite{Girvin99} in the electronic spin distribution as $\nu $
shifts slightly away from unity, and indicate the crucial importance of the
hyperfine interaction in controlling the nuclear spin dynamics \cite{VM88}, 
\cite{VM95}.

At filling factor $\nu =1$ the ground state of the 2D electron gas is
ferromagnetic even in the limit of zero Zeeman energy \cite{Sondhi93}.
Flipping nuclear spins in this state through the hyperfine interaction is
followed by the creation of spin excitons \cite{BIE81},\cite{KH84}. The
extremely long $T_{1}$ observed by Barrett \textit{et al} \cite
{Barret95,Tycko95} may be due to the energy gap existing in the exciton
spectrum (see below, however). In actual heterojunctions the gap is usually
much smaller than the theoretical value. It can be suppressed by the
combined effect of quantum confinement and external pressure \cite{Maude96}
or external electric fields \cite{Ivchenko97}. Furthermore, in 2D electron
gas under strong optical pumping the electronic Zeeman energy can be
strongly suppressed by the magnetic hyperfine field \cite{Kukushkin99}. In
both cases the effective g-factor is spatially inhomogeneous due to the
presence of long range impurity potential fluctuations \cite{VM95}.

Microscopic calculations, based on Hartree-Fock (HF) approximation for a
single, isolated Skyrmion \cite{FBCMKS97}, have found a family of low energy
excitations, with an approximately quadratic relation between the energy and
the number of flipped spins,$K$ , which can be associated with the kinetic
rotational energy of the entire spin texture. However, except for the
special case where $K$ is a half integer, the spectrum has an excitation
gap, which is some fraction of the large Coulomb energy scale. To account
for the observed enhancement in $1/T_{1}$, these authors have suggested \cite
{CMBFGS97} that at $\nu \neq 1$, where there is a finite density of
Skyrmions, the ground state is a Skyrme crystal, for which the spin waves
spectrum is gapless due to the breakdown of the global spin rotation
symmetry. The existence of such an ordered lattice at $\left| \delta \nu
\right| \lesssim 0.1$ , where $\delta \nu \equiv \nu -1$, is hard to
reconcile, however, with the radius, $R\lesssim 2l_{B}$, of the Skyrmions
reported in \cite{Barrett01}, as the average distance, $\left( 8/\left|
\delta \nu \right| \right) ^{1/2}$, between Skyrmions in this filling
factors region is larger than $9l_{B}$. The relatively large value of the
g-factor ( $g\approx -0.4$ ), characterizing these experiments, makes their
theoretical interpretation a difficult task.

In the present letter we propose to focus on different experimental
situations, like the ones mentioned above\cite{Maude96},\cite{Kukushkin99},
where the effective g-factor can become much smaller than the bulk GaAs
value (i.e. $\left| g\right| \approx 0.4$ ). We restrict the study to the
region $\delta \nu \equiv \left| \nu -1\right| \lesssim 0.05$ , so that the
spin rotation of individual Skyrmions with radii, $R\sim 7l_{B}$ ,
comparable to characteristic Skyrmion sizes found experimentally \cite
{Maude96},\cite{Kukushkin99}, may reasonably be regarded as independent. We
show that the excitation gap in the collective rotational spectrum of such a
single Skyrmion diminshes quite sharply when the Skyrmion radius increases,
so that for $R\sim 7l_{B}$ the spin excitation gap becomes comparable to the
nuclear Zeeman energy under consideration. Since the collective spin
rotation is coupled, through the hyperfine interaction, to nuclear spins,
the nuclear spin relaxation rate should be dramatically enhanced.

We start our analysis by considering the Hamiltonian for nuclear spins
interacting with 2D electron gas at $\nu =1$\ \ in MQW structure 
\begin{equation}
\widehat{H}=\widehat{H}_{n}+\widehat{H}_{e}+\widehat{H}_{en}
\label{Hamiltonian}
\end{equation}
where $\widehat{H}_{n}=-\hbar \gamma _{n}\sum_{j}\widehat{\mathbf{I}}%
_{j}\cdot \mathbf{B}_{0}$ is the nuclear Zeeman energy,$\ \widehat{H}_{e}=$
\ $-\hbar \gamma _{e}\int d^{2}r\widehat{\mathbf{S}}\left( \mathbf{r}\right)
\cdot \mathbf{B}_{0}+\widehat{H}_{ee}$ is the electronic Hamiltonian, with $%
\widehat{H}_{ee}$ the electron-electron interaction, and $\widehat{H}%
_{en}=A\sum_{j}\widehat{\mathbf{S}}\left( \mathbf{r}_{j}\right) \cdot 
\widehat{\mathbf{I}}_{j}$ is the Fermi contact hyperfine interaction between
the electron and nuclear spins. \ Here $\widehat{\mathbf{I}}_{j}$ is the
nuclear spin operator located at $\mathbf{r}_{j}$, $\widehat{\mathbf{S}}%
\left( \mathbf{r}\right) $ is the electronic spin density operator, $\mathbf{%
B}_{0}$ is the external magnetic field, which is perpendicular to the 2D
electron gas ( $\mathbf{B}_{0}=B_{0}\mathbf{z}$ ) , $\gamma _{n}=g_{n}\mu
_{n}/\hbar $ $\ $and $\gamma _{e}=g_{e}\mu _{B}/\hbar $ the nuclear and
electronic gyromagnetic ratios respectively, and $A=\frac{8\pi }{3}g_{n}\mu
_{n}g_{0}\mu _{B}\left| u_{0}\left( 0\right) \right| ^{2}$ is the Fermi
contact hyperfine coupling constant. In this expression $u_{0}\left(
0\right) $ is the periodic part of the Bloch wavefunction at the nucleus,
and $g_{0}$ is the g-factor of a free electron.

Within the HF approximation, and to leading order in a gradient expansion,
the electronic Hamiltonian $\widehat{H}_{e}$ can be written as a functional
of a unit vector field $\mathbf{n}\left( \mathbf{r}\right) $, 
\begin{equation}
\widehat{H}_{e}=\frac{\varepsilon _{C}}{32\pi }\int d^{2}r\left( \mathbf{%
\nabla \cdot n}\right) ^{2}-\frac{\varepsilon _{sp}}{4\pi l_{B}^{2}}\int
d^{2}r\left( \mathbf{z\cdot n}\right) -\frac{\varepsilon _{C}}{2}Q
\label{H_e}
\end{equation}
where $\varepsilon _{sp}=\left| g\right| \mu _{B}B_{0}$ is the Zeeman
splitting energy, $l_{B}=\sqrt{c\hbar /eB_{0}}$-the magnetic length, $%
\varepsilon _{C}=\sqrt{\frac{\pi }{2}}\frac{e^{2}}{\kappa l_{B}}$ is the
Coulomb energy, $\kappa $- the dielectric constant, and $Q=\frac{1}{4\pi }%
\int d^{2}r\left( \mathbf{n\cdot }\left[ \partial _{x}\mathbf{n\times }%
\partial _{y}\mathbf{n}\right] \right) $ is the Skyrmion winding number (or
topological charge). The unit vector field is proportional to the
expectation value of the electron spin density, i.e. $\mathbf{n}\left( 
\mathbf{r}\right) =4\pi \mathbf{S}\left( \mathbf{r}\right) $, with $\mathbf{S%
}\left( \mathbf{r}\right) \equiv \langle \widehat{\mathbf{S}}\left( \mathbf{r%
}\right) \rangle $. \ 

The dynamics of the Skyrmionic spin texture can be studied by examining the
corresponding effective Lagrangian. This should contain first order time
derivative to account for the spin wave excitations (or spin-excitons)
obtained within the microscopic HF scheme at $\nu =1$. \ It has the form $-%
\frac{1}{4\pi l_{B}^{2}}\int d^{2}r\left( \hbar \partial _{t}\mathbf{n}%
\right) \mathbf{\cdot A}\left( \mathbf{n}\right) $ , where $\mathbf{A}\left( 
\mathbf{n}\right) $ is a vector potential of a unit magnetic monopole
sitting at the origin in spin space \cite{Girvin99}. \ Since we are
interested here in a rigid rotation of the entire spin texture, we may
restrict the analysis, for the sake of simplicity, to the spatial region far
away from the Skyrmion core, where $\mathbf{n}$ is very close to $\mathbf{z}$
, so that $\mathbf{n}\approx \left(
n_{x},n_{y},1-n_{x}^{2}/2-n_{y}^{2}/2\right) $, and $\mathbf{A}\left( 
\mathbf{n}\right) \approx \frac{1}{2}\left( -n_{y},n_{x},0\right) $.
Separating the collective spin rotational motion, described by the variable $%
\varphi \left( t\right) $ , from all the other degrees of freedom in spin
space, by writing $\psi \left( \mathbf{r},t\right) \equiv n_{x}\left( 
\mathbf{r},t\right) +in_{y}\left( \mathbf{r},t\right) =\widetilde{\psi }%
\left( \mathbf{r},t\right) e^{i\varphi \left( t\right) }$,\ \ and then
adding \textit{a kinetic energy term}, $\frac{1}{2}m_{s}\psi ^{\ast }\psi
\left( \frac{\partial \varphi }{\partial t}\right) ^{2}$, \textit{associated
with the collective rotation angle}, $\varphi \left( t\right) $ , \textit{%
with} $m_{s}$ \textit{being a phenomenological mass, }the effective
Lagrangian density is written in the form: 
\begin{eqnarray}
&&\mathcal{L}\left\{ \widetilde{\psi },\widetilde{\psi }^{\ast };\varphi
\right\} =\frac{1}{4\pi l_{B}^{2}}\left\{ 
\begin{array}{c}
\varepsilon _{sp}\left( 1-\frac{1}{2}\widetilde{\psi }^{\ast }\widetilde{%
\psi }\right) + \\ 
\frac{i}{4}\left[ \widetilde{\psi }^{\ast }\left( \hbar \frac{\partial 
\widetilde{\psi }}{\partial t}\right) -\widetilde{\psi }\left( \hbar \frac{%
\partial \widetilde{\psi }^{\ast }}{\partial t}\right) \right] \\ 
-\frac{1}{2}\widetilde{\psi }^{\ast }\widetilde{\psi }\left( \hbar \frac{%
\partial \varphi }{\partial t}\right)
\end{array}
\right\}  \nonumber \\
&&-\frac{1}{32\pi }\varepsilon _{C}\left( \mathbf{\nabla }\widetilde{\psi }%
^{\ast }\cdot \mathbf{\nabla }\widetilde{\psi }\right) +\frac{1}{2}m_{s}%
\widetilde{\psi }^{\ast }\widetilde{\psi }\left( \frac{\partial \varphi }{%
\partial t}\right) ^{2}  \label{Lagr-mod}
\end{eqnarray}

Note that in deriving\ this expression all terms of order higher than $%
o\left( \left| \psi \right| ^{2}\right) $ are neglected. Furthermore, we
omit here the last term appearing on the RHS of Eq.(\ref{H_e}), which is
topological invariant (i.e. under any continuous deformation of $\mathbf{n}$%
), and so does not influence the equation of motion. \ \ Also note that Eq.(%
\ref{Lagr-mod}) is consistent with the effective Lagrangian derived from the
microscopic Hamiltonian in the HF approximation by Apel and Bychkov \cite
{AB97}.

It should be stressed that the degrees of freedom in spin space, described
by the field $\widetilde{\psi }\left( \mathbf{r},t\right) $ , are assumed to
be massless. This assumption is consistent with the lowest Landau level
approximation, usually exploited in this context. The Euler-Lagrange
equations, derived from the effective Lagrangian density, Eq.(\ref{Lagr-mod}%
), with respect to $\widetilde{\psi }^{\ast }$, and $\varphi $ , are: 
\begin{eqnarray}
&&-i\hbar \frac{\partial \widetilde{\psi }}{\partial t}+\varepsilon _{sp}%
\widetilde{\psi }-\frac{1}{4}\varepsilon _{C}l_{B}^{2}\nabla ^{2}\widetilde{%
\psi }+\left( \hbar \frac{\partial \varphi }{\partial t}\right) \widetilde{%
\psi }  \nonumber \\
&&-4\pi l_{B}^{2}m_{s}\left( \frac{\partial \varphi }{\partial t}\right) ^{2}%
\widetilde{\psi }=0  \label{class1} \\
&&\frac{d}{dt}\left[ \widetilde{\psi }^{\ast }\widetilde{\psi }-2\widetilde{m%
}_{s}\widetilde{\psi }^{\ast }\widetilde{\psi }\left( \frac{d\varphi }{dt}%
\right) \right] =0  \label{class2}
\end{eqnarray}

Since Eqs.(\ref{class1},\ref{class2}) describe the dynamics of spins in the
tail of the Skyrmion (where the spin polarization is nearly complete) the
spin-wave form: $\psi \left( \mathbf{r},t\right) =\widetilde{\psi }\left( 
\mathbf{r},t\right) e^{i\varphi _{0}}=\exp i\left( \mathbf{k}\cdot \mathbf{r}%
-\omega t+\varphi _{0}\right) $ , where $\varphi _{0}$ is a constant, should
be a solution. Indeed, substituting into Eqs.(\ref{class1},\ref{class2}),
the former equation becomes (the latter equation is trivially solved): $%
\left( -\hbar \omega +\varepsilon _{sp}+\frac{1}{4}\varepsilon
_{C}l_{B}^{2}k^{2}\right) \widetilde{\psi }=0$ , which is the well known
spin-wave dispersion relation in the long wave-length limit $\hbar \omega
=\varepsilon _{sp}+\frac{1}{4}\varepsilon _{C}l_{B}^{2}k^{2}$. \ 

The desired collective mode can be described by $\psi \left( \mathbf{r}%
,t\right) =\widetilde{\psi }_{0}\left( \mathbf{r}\right) e^{i\varphi \left(
t\right) }$ , where $\widetilde{\psi }_{0}\left( \mathbf{r}\right) $ is time
independent. It turns Eqs.(\ref{class1},\ref{class2}) respectively to 
\begin{eqnarray}
\left[ \varepsilon _{sp}\widetilde{\psi }_{0}-\frac{1}{4}\varepsilon
_{C}l_{B}^{2}\nabla ^{2}\widetilde{\psi }_{0}\right] &+&\widetilde{\psi }_{0}%
\Big[ \left( \hbar \frac{\partial \varphi }{\partial t}\right)-  \nonumber \\
4\pi l_{B}^{2}m_{s}\left( \frac{\partial \varphi }{\partial t}\right) ^{2}%
\Big] & =&0;\quad \left( \frac{\partial ^{2}\varphi }{\partial t^{2}}%
\right)=0
\end{eqnarray}

The first equation can be separated into purely spatial and temporal
equations: 
\[
\varepsilon _{sp}\widetilde{\psi }_{0}-\frac{1}{4}\varepsilon
_{C}l_{B}^{2}\nabla ^{2}\widetilde{\psi }_{0}=0,\;\ \ \left( \hbar \frac{%
\partial \varphi }{\partial t}\right) -4\pi l_{B}^{2}m_{s}\left( \frac{%
\partial \varphi }{\partial t}\right) ^{2}=0 
\]
with the nontrivial solution for the time dependent equation: 
\begin{equation}
\left( \frac{\partial \varphi }{\partial t}\right) =\frac{\hbar }{4\pi
l_{B}^{2}m_{s}}=\frac{e\left( \frac{1}{4\pi }\right) B_{0}}{m_{s}c}
\label{Omega_c}
\end{equation}

The solutions for the static, space dependent equation, $\widetilde{\psi }%
_{0}\left( r,\phi \right) \varpropto K_{m}\left( r/l_{sk}\right) e^{im\phi }$%
, with $K_{m}\left( x\right) $ a modified Bessel function of integer order, $%
m$, coincide with the asymptotic form of the static Skyrmion solutions with
winding number $m$ , found in Ref.(\cite{BMV96}). Here $l_{sk}$\ is the
length scale corresponding to the Skyrmion's tail, $l_{sk}^{-2}=4\varepsilon
_{sp}/\varepsilon _{C}l_{B}^{2}=2\sqrt{\frac{2}{\pi }}\left| g\right| 
\widetilde{a}/l_{B}^{3}$, and $\widetilde{a}=\kappa \hbar ^{2}/m_{0}e^{2}$
is the effective Bohr radius ($m_{0}$ being the free electron mass).

Eq.(\ref{Omega_c}) yields an effective Larmor frequency for precession of
the entire spin texture in a magnetic field $B_{s}\equiv \left( \frac{1}{%
4\pi }\right) B_{0}$. The remarkable feature of this result is that $B_{s}$
is , up to a constant, identical to the external magnetic field $B_{0}$.

The collective mass $m_{s}$ remains unknown within the phenomenological
approach described above. It may be determined from the variational Skyrmion
energy, derived in \cite{BMV96}, by exploiting the connection \cite{Girvin99}
between the total number, $K$ , of flipped spins in the Skyrmion and the
eigenvalues of $\widehat{L}_{z}$-the canonical angular momentum conjugate to 
$\varphi $. The variational Skyrmion energy was calculated in the HF
approximation near filling factor $\nu =1$\cite{BMV96} by adding to the
Hamiltonian $\widehat{H}_{e}$ in Eq.(\ref{Hamiltonian}), consisting of the
Zeeman energy and the leading exchange energy term in the gradient
expansion, the Coulomb self energy repulsion. The latter represents, in
addition to the leading, topological invariant term, higher order
corrections in the gradient expansion, resulting in the following expression
for the total Skyrmion energy: 
\begin{equation}
E_{tot}\left( R\right) =\frac{3\pi ^{2}e^{2}}{2^{6}\kappa R}+\frac{1}{4}%
\varepsilon _{C}\left( \frac{R}{l_{sk}}\right) ^{2}\ln \left( \frac{2l_{sk}}{%
R}\right)  \label{Etot}
\end{equation}
where $R$ is a variational parameter describing the Skyrmion core radius. \
Identifying $L_{z}$ with the number of flipped spins in the Skyrmion\cite
{Girvin99}, the Zeeman energy is $\Delta E_{Z}=\varepsilon _{sp}\widetilde{L}%
_{z}=\frac{1}{4}\varepsilon _{C}\left( \frac{R}{l_{sk}}\right) ^{2}\ln
\left( \frac{2l_{sk}}{\sqrt{\bar{e}}R}\right) $ , where, $\widetilde{L}%
_{z}\equiv L_{z}/\hbar $ , and $\bar{e}$ stands for the natural logarithm
base, so that $\widetilde{L}_{z}=\left( \frac{R}{l_{B}}\right) ^{2}\ln
\left( \frac{2l_{sk}}{\sqrt{\bar{e}}R}\right) $. Minimization with respect
to $R$ yields for the equilibrium core radius $\frac{3\pi ^{2}e^{2}}{%
2^{6}\kappa R_{eq}^{3}}=\varepsilon _{sp}\left( \frac{2}{l_{B}^{2}}\right)
\ln \left( \frac{2l_{sk}}{R_{eq}}\right) $, whereas $\left[ \frac{\partial
^{2}}{\partial R^{2}}E_{tot}\right] _{eq}\approx \left( \frac{6\varepsilon
_{sp}}{l_{B}^{2}}\right) \ln \left( \frac{2l_{sk}}{R}\right) $, so that: 
\begin{equation}
U=\left[ \frac{\partial ^{2}}{\partial \widetilde{L}_{z}^{2}}E_{tot}\right]
_{eq}\approx \varepsilon _{sp}\left( \frac{l_{B}}{R_{eq}}\right) ^{2}\frac{%
3\ln \left( \frac{2l_{sk}}{R}\right) }{2\ln ^{2}\left( \frac{2l_{sk}}{\sqrt{%
\bar{e}}R}\right) }  \label{Ueq}
\end{equation}

Expanding the energy, Eq.(\ref{Etot}), up to second order in $\widetilde{L}%
_{z}$ about its equilibrium value, $K$, that is writing $E_{tot}\left( 
\widetilde{L}_{z}\right) =E_{tot}\left( K\right) +\frac{1}{2}U\left( 
\widetilde{L}_{z}-K\right) ^{2}+...$,\ the second term on the RHS
corresponds to the 'classical' rotational energy of the entire spin texture
about its symmetry axis. At the classical level any deviation of $\widetilde{%
L}_{z}$ from its equilibrium value $K$ corresponds to a continuous spatial
deformation of the Skyrmion with respect to its equilibrium configuration,
thus conserving its topological charge,$Q$, but increasing the Skyrmion
energy with respect to its equilibrium value. The collective rotation of the
Skyrmion in spin space is therefore reflected as a radial expansion or
contraction in orbital space. Quantization of this rotational motion can be
achieved by replacing $L_{z}\rightarrow \frac{\hbar }{i}\frac{\partial }{%
\partial \varphi }$ , which yields $\widehat{H}_{rot}=\frac{1}{2}U\left( 
\frac{1}{i}\frac{\partial }{\partial \varphi }-K\right) ^{2}$. \ Now, the
rotational energy \cite{Girvin99}, $E_{rot}=\frac{\hbar ^{2}}{2U}\left( 
\frac{d\varphi }{dt}\right) ^{2}$, may be equated to\ the lowest eigenvalue
of the rotational Hamiltonian $\widehat{H}_{rot}$, namely $\sim \frac{1}{2}U$
, to find the angular velocity: 
\begin{equation}
\left( \frac{d\varphi }{dt}\right) \sim U/\hbar =\frac{eB_{0}}{M_{s}c},\quad
M_{s}=4\left( \frac{\widetilde{a}}{l_{B}}\right) \left| \widetilde{g}\right|
^{-5/3}m_{0}  \label{Omega_cHF}
\end{equation}
and $\widetilde{g}\equiv g\left( \frac{\widetilde{a}}{l_{B}}\right) $. This
remarkable result is consistent with the Larmor frequency found in Eq.(\ref
{Omega_c}), provided the phenomenological mass is $m_{s}=M_{s}/4\pi $. The
resulting effective mass, Eq.(\ref{Omega_cHF}), diverges with vanishing
g-factor like $\left| \widetilde{g}\right| ^{-5/3}$, reflecting the
macroscopic inertial mass associated with the collective rotation of a
Skyrmion.\ 

As indicated above, local g-factors should fluctuate significantly in the
space of the QW, due e.g. to lattice strains, and long range fluctuating
electric field, which always exists in the QW as a result of the ionized
impurities located in the barriers. In fact for a QW width $l\sim 70\AA $ 
\cite{Maude96}, and typical electric field of about $10^{-3}V/\AA $, the
corresponding fluctuation in the g-factor can be of the order of the bulk
g-factor itself \cite{Ivchenko97}, so that local values can become very
small. The size of the corresponding large Skyrmions is expected, however,
to be limited by disorder potential \cite{Lilliehook97}, \cite{Nederveen99}, 
\cite{Barrett01}. A lower bound on the values of $\left| g\right| $ , below
which the Skyrmion radius does not change, can be estimated from the
experimental data reported in Ref.(\cite{Maude96}). This yields $g\sim 0.03$
( or $\widetilde{g}\sim 0.002$ ) for which the corresponding mass ratio is $%
M_{s}/m_{0}\sim 10^{4}$, and the collective Larmor frequency, $\omega _{sk}=%
\frac{eB_{0}}{M_{s}c}$, is comparable to the nuclear $^{71}$Ga Zeeman
frequency, $\omega _{n}\approx 10^{-4}\gamma _{e}B_{0}$. It should be
stressed, however, that despite the huge enhancement of $M_{s}$, the radius, 
$R\approx 7l_{B}$, of such a large spin texture, remains smaller than half
of the average distance between neighboring Skyrmions, as long as $\left|
\delta \nu \right| \leq 0.05$. Thus, within this filling factor region our
picture of independently rotating Skyrmions in spin space should be valid.

Our theoretical tools are not yet sufficiently developed, however, to enable
us sufficiently accurate\ evaluation of the desired relaxation rates. The
best we can do at present is to set some reasonable bounds on these rates.
We follow the theory developed in Refs. \cite{AB01}, \cite{MBVW01} to
estimate the decay of the deviation, $\delta I_{z}(\mathbf{r},t)$, of the
nuclear spin polarization from its equilibrium value, i.e. $\delta I_{z}(%
\mathbf{r},t)=\delta I_{z}(\mathbf{r},0)e^{-\Gamma \left( \mathbf{r}%
,t\right) }$, where $\Gamma \left( \mathbf{r},t\right) =\func{Re}%
\int_{0}^{t}dt^{\prime }\xi \left( \mathbf{r},t^{\prime }\right) $, $\xi
\left( \mathbf{r},t\right) =\frac{\alpha ^{2}}{2}\int_{0}^{t}d\tau
e^{i\omega _{n}(\tau -t)}\left\langle \left\{ \widehat{S}_{+}(\mathbf{r},t),%
\widehat{S}_{-}(\mathbf{r},\tau )\right\} \right\rangle $, and the average
is performed over the electronic states. Restricting the analysis to
positions $\mathbf{r}$ far away from the Skyrmion core, where $S_{+}(\mathbf{%
r},t)\approx \frac{1}{4\pi }\widetilde{\psi }_{0}\left( \mathbf{r}\right)
e^{i\varphi \left( t\right) }$, we may rewrite $\xi \left( \mathbf{r}%
,t\right) \approx 2\left( \frac{\alpha \left| \widetilde{\psi }_{0}\left( 
\mathbf{r}\right) \right| }{8\pi }\right) ^{2}\int_{0}^{t}d\tau e^{i\omega
_{n}(\tau -t)}\left\langle \left\{ e^{i\widehat{\varphi }\left( t\right)
},e^{-i\widehat{\varphi }\left( \tau \right) }\right\} \right\rangle $,
where $e^{i\widehat{\varphi }\left( t\right) }\equiv e^{it\widehat{H}%
_{rot}/\hbar }e^{i\varphi }e^{-it\widehat{H}_{rot}/\hbar }$. \ A
straightforward algebra yields: $\Gamma \left( \mathbf{r},t\right) =\left( 
\frac{\alpha \left| \widetilde{\psi }_{0}\left( \mathbf{r}\right) \right| }{%
4\pi }\right) ^{2}\frac{1-\cos \left[ \left( U\delta K/\hbar -\omega
_{n}\right) t\right] }{\left( U\delta K/\hbar -\omega _{n}\right) ^{2}}$,
where $\delta K\equiv \left[ K\right] -K$ , and $\left[ K\right] $ is the
half integer closest to $K$. \ This expression, which describes a highly
idealized situation, may be used to estimate a lower bound for $T_{1}$ by
considering the limit when the excitation gap vanishes, i.e. when $\left(
U\delta K/\hbar -\omega _{n}\right) \rightarrow 0$. The corresponding
nuclear spin relaxation is Gaussian, $\delta I_{z}(\mathbf{r},t)\sim $ \ $%
e^{-\left( \alpha \left| \widetilde{\psi }_{0}\left( \mathbf{r}\right)
\right| /4\pi \right) ^{2}t^{2}}$, with characteristic local relaxation time 
$T_{1}^{sk}\left( \mathbf{r}\right) \sim 4\pi /\alpha \left| \widetilde{\psi 
}_{0}\left( \mathbf{r}\right) \right| $, determined by the local transverse
spin density. Typical values are therefore of the order of$\ 1/K_{S}$, which
is roughly $10^{-3}\sec $ for GaAs MQW. This extremely small result (as
compared to typical values, $T_{1}\sim 10^{3}\sec $ , observed at $\nu =1$ )
indicates the great sensitivity of the nuclear relaxation time $T_{1}$ to
the filling factor in the close vicinity of $\nu =1$. \ 

It is interesting to make a similar lower bound estimate of the
corresponding relaxation time at $\nu =1$, where the number of Skyrmions
vanishes and the nuclear spin dynamics is controlled by the coupling to the
spin waves. In this case \cite{MBVW01}: $\Gamma \left( \mathbf{r},t\right)
=\Gamma (t)=2\left( hK_{S}\right) ^{2}\int_{0}^{\infty }\widetilde{k}d%
\widetilde{k}e^{-\widetilde{k}^{2}/2}\frac{1-\cos ([\varepsilon _{ex}(%
\widetilde{k})/\hbar -\varpi ]t)}{[\varepsilon _{ex}(\widetilde{k})-\hbar
\varpi ]^{2}}$, where $\varepsilon _{ex}(\widetilde{k})\approx \varepsilon
_{sp}+\frac{1}{4}\varepsilon _{C}\widetilde{k}^{2}$ , for $\widetilde{k}%
=kl_{H}\ll 1$. Thus in the limit $\varepsilon _{ex}(\widetilde{k})-\hbar
\varpi \rightarrow 0$ , we find for $t\gg \hbar /\varepsilon _{C}$ : $\Gamma
(t)=\left( 2hK_{S}\right) ^{2}\int_{0}^{\infty }\widetilde{k}d\widetilde{k}%
e^{-\widetilde{k}^{2}/2}\frac{\sin ^{2}(\varepsilon _{C}\widetilde{k}%
^{2}t/8\hbar )}{(\varepsilon _{C}\widetilde{k}^{2}/4)^{2}}\rightarrow \left( 
\frac{2\left( 2\pi \right) ^{2}K_{S}^{2}}{\varepsilon _{C}/h}\right) t$ , \
so that the relaxation of the nuclear spin polarization is a simple
exponential, with a characteristic relaxation time $T_{1}^{exc}=\frac{1}{%
8\pi ^{2}}\left( \frac{\varepsilon _{C}/h}{K_{S}}\right) \left( \frac{1}{%
K_{S}}\right) $. \ For GaAs MQW this expression yields $T_{1}^{exc}\sim
4\times \allowbreak 10^{2}$ sec, which is of the same order of magnitude as
the experimentally measured $T_{1}$ at $\nu =1$. The very long $T_{1}$
obtained in this case is due to the large (Coulomb) energy scale of the
spin-waves excitations. In contrast, the energy scale of the collective spin
rotations of all possible Skyrmions with different g-factors is always a
small fraction of the electronic Zeeman energy, so that the corresponding $%
T_{1}$ is by several orders of magnitude shorter than $T_{1}^{exc}$.

In conclusion, we have shown that for a QH ferromagnet, realized in QW with
sufficiently suppressed effective g-factor \cite{Maude96},\cite{Kukushkin99}%
, one expects dramatic enhancement of the nuclear spin-lattice relaxation
rate, while shifting the filling factor slightly away from $\nu =1$. This
effect is associated with nearly gapless excitation of almost free rotations
in spin space of individual Skyrmions.

We thank Sean Barret for illuminating discussions. This research was
supported by the S. Langberg Nuclear Research Fund at the Technion, and by
European Commission grants: HPRI-CT-1999-400013 and INTAS-2001-0791. Yu.
A.B. acknowledges the support from Grant RFFI- 03-02-16012.

\end{document}